\documentclass[twocolumn,showpacs,amsmath,amssymb,pra]{revtex4}
\usepackage{textcomp}
\usepackage{amsmath}
\usepackage{amssymb}
\usepackage{graphicx}
\usepackage{epstopdf}
\usepackage{bbm}
\usepackage[toc,page,title,titletoc,header]{appendix}
\usepackage{dcolumn}\usepackage{bm}
\newcommand{\be}{\begin{equation}}
\newcommand{\ee}{\end{equation}}
\newcommand{\bey}{\begin{eqnarray}}
\newcommand{\eey}{\end{eqnarray}}
\newcommand{\bw}{\begin{widetext}}
\newcommand{\ew}{\end{widetext}}
\newcommand{\ww}{\widetilde}

\newcommand{\ov}{\overline}

\newcommand{\ra}{\rangle}
\newcommand{\la}{\langle}

\newcommand{\ba}{\begin{array}}
\newcommand{\ea}{\end{array}}
\newcommand{\bi}{\begin{itemize}}
\newcommand{\ei}{\end{itemize}}
\newcommand{\bem}{\begin{enumerate}}
\newcommand{\eem}{\end{enumerate}}

\makeatother

\begin{document}

\title{Statistical distribution of components of energy eigenfunctions:
from nearly-integrable to chaotic }

 \author{Jiaozi Wang and Wen-ge Wang
 \footnote{ Email address: wgwang@ustc.edu.cn}
  }

\affiliation{Department of Modern Physics, University of Science and Technology of China,
Hefei, 230026, China}

\date{\today}

\begin{abstract}
  We study the statistical distribution of components in the non-perturbative parts
  of energy eigenfunctions (EFs), in which main bodies of the EFs lie.
  Our numerical simulations in five models show that
  deviation of the distribution from the prediction of random matrix theory (RMT)
  is useful in characterizing the process from nearly-integrable to chaotic,
  in a way somewhat similar to the nearest-level-spacing distribution.
  But, the statistics of EFs reveals some more properties, as described below.
  (i) In the process of approaching quantum chaos, the distribution of
  components shows a delay feature compared with the nearest-level-spacing distribution
  in most of the models studied.
  (ii) In the quantum chaotic regime,
  the distribution of components always shows small but notable deviation from the prediction of RMT
  in models possessing classical counterparts, while, the deviation can be
  almost negligible in models not possessing classical counterparts.
  (iii) In models whose Hamiltonian matrices possess a clear band structure,
  tails of EFs show statistical behaviors obviously different from those
  in the main bodies, while, the difference
  is smaller for Hamiltonian matrices without a clear band structure.
\end{abstract}


\maketitle

\section{Introduction}\label{sect-intro}

 Statistical properties of energy eigenfunctions (EFs) of quantum chaotic systems
 have been studied extensively in the past several decades
 (see, e.g., Refs.\cite{Haake,CC94book,Stockmann,Berry77,Meredith98,Buch82,Sr96,
 Iz96,Connor87,Sr98,Backer02,Urb03,
 scanz05,Mirlin00,Mirlin02,KpHl,Kp05,Bies01,Falko96,Prosen03,Lewf95,Prig95,pre-98,
 Gnutzmann10,Anh11,stockmann09,Kaplan09,pre02-LMG}).
 These properties are of interest in many fields in physics, e.g., in the study of
 chaotic quantum dots \cite{qd1,qd2}, in that of optical, elastomechanical,
 and microwave resonators \cite{opt1,opt3,mech1,micw2,micw3,micw4,ccchen},
 and in the study of decay and fluctuations of heavy nuclei \cite{nuc1,nuc2,nuc3}.
 They are also of relevance to thermalization
 \cite{Izlev11,Haake12,Deutsch91,Sr94,Rigol08,Rigol12,Robin14,Gemmer15},
 a topic which has attracted renewed interest in recent years.

 Compared with statistical properties of the spectra of quantum chaotic systems,
 which can be described by the random matrix theory (RMT)
 \cite{Haake,CC94book,Stockmann,Bohigas,berry85,Sieber-Richter00,Gutzbook,Haake04},
 those of EFs are much more complex and satisfactory answers to many important
 questions are still lacking.
 For example, to what extent could components in what parts of EFs
 of quantum chaotic systems have a distribution close to the prediction of RMT?
 To what extent could deviation of the distribution of components from the
 prediction of RMT be used to characterize a transition process from integrable to chaotic?

 In the study of statistics of EFs, several problems are inevitably faced.
 One is the basis-dependence problem ---
 clearly, EFs may have different statistical properties in different bases.
 This problem is not serious in many models of practical interest,
 because in these models certain specific bases may be of interest due to physical reasons,
 e.g., the configuration space or certain unperturbed basis.
 For EFs in the configuration space, Berry's conjecture
 supplies a basic framework for investigation \cite{Berry77},
  meanwhile, specific dynamics of the underlying classical
 system may induce certain modifications \cite{KpHl,Bies01,Sr98,Backer02,Urb03}.
 However, for EFs in unperturbed bases, not as much is known for
 their statistical properties.

 Another problem is related to the location of  main bodies of EFs.
 Indeed, long-tails of EFs with fast decay should show statistical behaviors
 different from those of the main bodies, and this may lead to deviation
 from the prediction of RMT (see, e.g., Ref.\cite{Meredith98}).
 In the configuration space, the location can be approximately given by
 the classically-allowed region \cite{Berry77,Gutzbook,LL-SP}.
 However, the situation is not so clear in the case of unperturbed basis.

 A promising method of locating main bodies of EFs is given by
 a generalized Brillouin-Wigner perturbation theory (GBWPT) \cite{pre-98}.
 The GBWPT shows that an EF can be divided into a non-perturbative (NPT) part
 and a perturbative (PT) part, with the PT part expanded in a convergent
 perturbation expansion by making use of the NPT part.
 For a Hamiltonian matrix with a band structure,
 PT parts of its EFs show an exponential-type decay and,
 hence, main bodies of the EFs should lie in the NPT parts \cite{pre00}.
 In fact, in the Lipkin-Meshkov-Glick (LMG) model \cite{LMG} with a banded Hamiltonian matrix,
 numerical simulations show that, for those EFs that are delocalized in their NPT parts,
 the distribution of components in their NPT parts is quite close to the prediction
 of RMT \cite{pre02-LMG}.

 In this paper, we carry out a systematic study of the statistical distribution
 of components in NPT parts of EFs.
 We study all EFs in the middle energy region in five models with different features,
 some possessing classical counterparts and some not,
 some having banded Hamiltonian matrices and some not.
 As well known, deviation of the nearest-level-spacing distribution from the prediction of RMT
 can be used as a measure of the ``distance to quantum chaos'' \cite{Haake}.
 We show that deviation of the above-mentioned distribution
 from the prediction of RMT is also useful in characterizing
 the process from integrable to chaotic and, further, it may reveal more
 than that supplied by the former deviation.

 The paper is organised as follows. 
 In Sec.\ref{sect-GBWPT}, we briefly recall basic results of the GBWPT, in particular,
 the division of PT and NPT parts of EFs.
 In Sec.\ref{sect-Model}, we introduce the models to be employed.
 Numerical results obtained in models possessing classical counterparts are
 discussed in Sec.\ref{sect-m-cc}.
 Those in models without classical counterparts are given in Sec.\ref{sect-m-non-cc}.
 Finally, summaries and discussions are given in Sec.\ref{sect-Conclusion}.

\section{The GBWPT and the models employed}

 In this section, we first recall basic contents of the GBWPT,
 then, discuss the models to be employed in our numerical simulations.

\subsection{Generalized Brillouin-Wigner perturbation theory}
\label{sect-GBWPT}

 Consider a Hamiltonian of the form
\begin{equation}\label{H}
 H(\lambda)=H_{0}+\lambda V,
\end{equation}
 where $H_{0}$ is an unperturbed Hamiltonian and $\lambda V$ represents a generic perturbation
 with a control parameter $\lambda$.
 Eigenstates of $H(\lambda)$ and $H_{0}$ are denoted by $|\alpha\rangle$ and $|k\rangle$,
 respectively,
\begin{equation}
H(\lambda)|\alpha\rangle=E_{\alpha}(\lambda)|\alpha\rangle,
\quad H_{0}|k\rangle=E_{k}^{0}|k\rangle,
\end{equation}
 with labels $\alpha$ and $k$ in energy order.
 Components of the EFs are denoted by $C_{\alpha k} = \la k|\alpha\ra$
 and the dimension of the Hilbert space is denoted by $d_H$.

 In the GBWPT, for each perturbed
 state $|\alpha\rangle$, the set of unperturbed states $|k\rangle$
 is divided into two substes, denoted by $S_{\alpha}$ and $\overline{S}_{\alpha}$.
 The related projection operators,
\begin{equation}
P_{S_{\alpha}}=\sum\limits _{|k\rangle\in S_{\alpha}}|k\rangle\langle{k}|,
\ Q_{\overline{S}_{\alpha}}
=\sum\limits _{|k\rangle\in\overline{S}_{\alpha}}|k\rangle\langle k|=1-P_{S_{\alpha}},
\end{equation}
 divide the perturbed state into two parts,
$|\alpha_{s}\rangle\equiv{P_{S_{\alpha}}|\alpha\rangle}$,
$|\alpha_{\overline{s}}\rangle\equiv Q_{\overline{S}_{\alpha}}|\alpha\rangle$.
 As shown in Ref.\cite{pre-98}, if this division satisfies the following condition, namely
\begin{equation}
 \lim _{n \to \infty }  \langle \phi |(T_{\alpha }^{\dagger})^n
T_{\alpha }^n |\phi  \rangle =0 \quad \forall |\phi\ra,  \label{conv}
\end{equation}
 where
\begin{equation}\label{T-alpha}
T_{\alpha}=\frac{1}{E_{\alpha}-H_{0}}Q_{\overline{S}_{\alpha}}\lambda{V},
\end{equation}
 then, making use of the part $|\alpha_{s}\rangle$,
the other part $|\alpha_{\overline{s}}\rangle$ can be expanded
in a convergent perturbation expansion, i.e.,
\begin{equation}\label{alpha-ovs}
|\alpha_{\overline{s}}\rangle=T_{\alpha}|\alpha_{s}\rangle+T_{\alpha}^{2}|\alpha_{s}\rangle
+\cdots+T_{\alpha}^{n}|\alpha_{s}\rangle+\cdots.
\end{equation}

 Let us consider an operator $W_{\alpha }$ in the subspace spanned by unperturbed states
 $|k\ra \in {\overline S}_{\alpha } $,
\be W_{\alpha } := Q_{{\overline S}_{\alpha }} V \frac{1}{E_{\alpha }-H^0} Q_{{\overline S}_{\alpha }},
\label{U} \ee
 and use $|\nu\rangle $ and $w_{\nu }$ to denote its eigenvectors
 and eigenvalues, $W_{\alpha } |\nu\rangle = w_{\nu } |\nu\rangle $,
 where for brevity we omit the subscript
 $\alpha$ for $|\nu\rangle $ and $w_{\nu }$.
 It is easy to verify that the condition (\ref{conv})  is equivalent to the requirement that
\begin{equation} \label{inequ}
 |\lambda w_{\nu }| < 1 \quad \forall |\nu\rangle.
\end{equation}

 In a quantum chaotic system $H(\lambda)$,
 all good quantum numbers of the unperturbed system $H_0$, except that related to the energy,
 are destroyed.
 For this reason, we consider $S_\alpha$ of the form
\begin{equation}\label{S-alpha}
 S_\alpha = \{ |k\ra : k_1 \le k \le k_2 \}.
\end{equation}
 We call the smallest set $S_{\alpha}$ satisfying the condition (\ref{conv})
 the \emph{non-perturbative (NPT) region} of the state $|\alpha\ra$, and
 the related set $\ov S_{\alpha}$ the \emph{perturbative (PT) region}.
 Correspondingly, the state $|\alpha\ra$ is divided into a NPT part
 $|\alpha_{s}\rangle$ and a PT part $|\alpha_{\overline{s}}\rangle$.
 Clearly, the NPT region of $|\alpha\ra$ has the smallest value of $(k_2-k_1)$.

 For sufficiently small $\lambda$ and for $E_\alpha$ not very close to
 any of the unperturbed eigenenergies,
 the condition (\ref{conv}) can be satisfied with $S_{\alpha} = \{ |k_0\ra \}$,
 where $|k_0\ra$ is the unperturbed state whose energy is the closest to $E_\alpha$.
 In this case, $k_1=k_2 = k_0$.
 With increasing perturbation strength $\lambda$, usually the width the NPT region increases.

 As an application of the GBWPT, let us discuss a Hamiltonian whose matrix has
 a band structure in the unperturbed basis.
 Expanding the state vector $Q_{{\overline S}_{\alpha }}\lambda V |\alpha _s\rangle $
 in the basis $|\nu\rangle $, one has
\begin{equation}\label{Qas-expan}
 Q_{{\overline S}_{\alpha }}\lambda V |\alpha _s\rangle = \sum_{\nu } d_{\nu } |\nu\rangle .
\end{equation}
 Substituting Eq.(\ref{T-alpha}) and Eq.(\ref{Qas-expan}) into Eq.(\ref{alpha-ovs}),
 after simple derivation, it is found that, for each unperturbed state $|j\rangle $
 in the set ${\overline S}_{\alpha }$,
 the component $C_{\alpha j}=\langle j|\alpha \rangle $ is written as
\begin{equation}
 C_{\alpha j} = \frac 1{E_{\alpha }-E^0_j} \sum_{\nu }
\left [ \frac {d_{\nu }}{1- \lambda w_{\nu}} \langle j|\nu\rangle \right ]
\left ( \lambda w_{\nu } \right )^{m-1} , \label{CU}
\end{equation}
 where $m$ is the smallest positive integer for
 $\langle j |(Q_{{\overline S}_{\alpha }}V)^m |\alpha _s\rangle $ not equal to zero,
 i.e., the smallest number of steps of coupling from $|j\ra$ to
 $|\alpha_s\ra$ through the perturbation $V$ \cite{pre00}.
 Let us use $b$ to denote the band width of the Hamiltonian matrix,
 thus, $\la k| V|k'\ra =0$ if $|k-k'| >b$.
 For $j>k_2$, one has $m \ge \frac{1}{b}(j-k_2)$, then,
 since $|\lambda w_{\nu }| < 1$, Eq.(\ref{CU}) shows that the EF has
 an exponential-type decay with increasing $j$.
 Similarly, the EF has an exponential-type decay with decreasing $j$ for $j<k_1$.

 Special attention should be paid to the two regions $[k_1-b,k_1]$ and $[k_2,k_2+b]$,
 for which $m=1$.
 According to Eq.(\ref{CU}), the exponential-type decay does not appear in these two regions.
 We call these two regions the \emph{shoulders} of the NPT region.
 Usually, the main body of the EF should lie within the region $[k_1-b,k_2+b]$, namely,
 within the NPT-plus-shoulder region.

\subsection{Models Employed}
\label{sect-Model}

 We employ five models in our numerical simulations,
 all exhibiting quantum chaos in certain parameter regions.
 Here, quantum chaos is defined by closeness of the statistics of the spectrum,
 particularly, the nearest-level-spacing distribution to the prediction of RMT \cite{Haake}.

 The first two models,
 a three-orbital LMG model \cite{LMG} and a single-mode Dicke model \cite{Dicke},
 possess classical counterparts.
 While, the rest three models, the Wigner-band random-matrix (WBRM) model \cite{WBRM},
 a defect XXZ chain \cite{d-XXZ}, and a defect Ising chain \cite{d-Ising},
 do not have any classical counterpart.
 Hamiltonian matrices in the LMG, the Dicke, and the WBRM models
 have a clear band structure in the considered unperturbed bases,
 while, those in the rest two models do not have a clear band structure.
 Below, we give brief descriptions of the models.

 The three-orbital LMG model is composed of $\Omega$ fermionic particles,
 occupying three energy levels labeled by $r=0,1,2$,  each possessing $\Omega$-degeneracy.
 Here, we are interested in the collective motion of this model.
 We use $\epsilon_{r}$ to denote the energy of the $r$-th level
 and, for brevity, we set $\epsilon_{0}=0$.
 The Hamiltonian of the model, in the form of Eq.(\ref{H}), is given by
\begin{equation}
 H_{0}=\epsilon_{1}K_{11}+\epsilon_{2}K_{22}, \quad V=\sum_{t=1}^{4}\mu_{t}V^{(t)}.
\end{equation}
 Here, $K_{rr} $ represents the particle number operator for the orbital $r$ and
\begin{eqnarray}
V^{(1)}=K_{10}K_{10}+K_{01}K_{01},\ V^{(2)}=K_{20}K_{20}+K_{02}K_{02},\nonumber \\
 V^{(3)}=K_{21}K_{20}+K_{02}K_{12},\ V^{(4)}=K_{12}K_{10}+K_{01}K_{21}, \ \
\end{eqnarray}
 where $K_{rs}$ with $r\neq s$ indicate particle raising and lowering operators.
 (See Refs.\cite{Meredith98,pre-98} for detailed properties of the operators $K_{rs}$
 and of the Hamiltonian matrix.)
 In our numerical simulations, the particle number is set at $\Omega=40$,
 as a result, the dimension of the Hilbert space is $d_H =861$.
 Other parameters are $\epsilon_{1}= 1.1, \epsilon_{2} = 1.61,
\mu_{1} = 0.039, \mu_{2} = 0.044, \mu_{3} = 0.048$, and $\mu_{4} = 0.041$.
 With these parameters, the nearest-level-spacing distribution
 is quite close to the Wigner distribution at $\lambda$ about $1$.

 The single-mode Dicke model
 describes the interaction between a single bosonic mode and a collection of $N$ two-level atoms.
 The system can be described in terms of a collective operator ${\bf \hat{J}}$ for the $N$
 atoms, with
 \be
 \hat{J}_{z} \equiv \sum_{i=1}^{N} \hat{s}_{z}^{(i)},\ \ \hat{J}_{\pm} \equiv \sum_{i=1}^{N}
 \hat{s}_{\pm}^{(i)},
 \ee
 where $\hat{s}_{x (y,z)}^{(i)}$ represent Pauli matrices divided by $2$ for the $i$-th atom.
 The operators $J_{z,\pm}$ obey the usual commutation rules for the angular momentum,
\be
[J_z , J_{\pm}]=\pm J_{\pm},\ \ [J_+,J_-]=2J_z.
\ee
 The unperturbed Hamiltonian and the perturbation are written as \cite{Emary03}
 \bey
H_0 & = & \omega_0 J_z +\omega a^{\dagger}a ,\   \nonumber \\
V & = & \frac{1}{\sqrt{2j}}(a^{\dagger}+a)(J_+ +J_-),
\eey
 where $j=N/2$.
 In numerical simulations, we take $N=40$ and truncate the particle number of the
 bosonic mode at $40$.
 We consider the resonance condition with $\omega_0 =\omega=1$.
 The Hilbert space is spanned by eigenstates of $J_z$, denoted by $|m\rangle$
 with $m=-j,-j+1,\cdots,j-1,j$.

 In the WBRM model,
 the unperturbed Hamiltonian $H_0$ takes a diagonal form with $E^0_i=i$  ($i=1 \cdots ,d_H$).
 The elements $v_{ij}$ of the perturbation $V$ are random numbers with Gaussian distribution for
 $1\leqslant |i-j| \leqslant b$ $(\langle v_{ij} \rangle =0, \langle v^2_{ij} \rangle=1$)
 and are zero otherwise.
 Thus, the Hamiltonian matrix has a band structure with a bandwidth $b$.
 This model exhibits a phenomenon called ``localization in the energy shell''
 at large $\lambda$ \cite{{CCGI96}}.

 The defect XXZ model is a modified XXZ model, with a magnetic field
 applied to one of the $N$ spins.
 We use the free boundary condition.
 The unperturbed Hamiltonian and the perturbation are written as
\bey
H_0  =  \mu_{d} \sigma^{5}_z+\sum_{i=1} ^{N-1}\mu \sigma^i _z \sigma^{i+1} _z,
\nonumber \\
V=\sum_{i=1} ^{N-1}  \sigma^i _x \sigma^{i+1} _x + \sigma^i _y \sigma^{i+1} _y.
\eey
 The system is integrable without the additional magnetic field applied at the
 $N/2$-th spin,
 but, it can be a quantum chaotic system when the additional magnetic field is sufficiently strong.
 The total Hamiltonian $H$ is commutable with $S_z$, the $z$-component of the total spin,
 and we use the subspace of $S_z = -2$ in our numerical study.
 Other parameters used in this model are $N=12$, $\mu_d =1.11$, and $\mu = 0.5$.

 The defect Ising model has a defect term similar to that
 in the defect XXZ model discussed above and,
 similarly, it can be a quantum chaotic system when the additional field is sufficiently strong.
 In this model,
\bey
H_0 = \mu_{d} \sigma^{5}_z+\sum^N_i J_z \sigma^i_z \sigma^{i+1}_z,
\ V= \sum_{i=1}^{N-1} { \sigma^i _x }. \ \ \
\eey
 Parameters used are $N=10$, $\mu_d = 1.11$, $J_z=1$, and $\mu_z=0.4$.

 For the sake of convenience when comparing results obtained in different models,
 we rescale the parameter $\lambda \to \lambda' = \lambda/a$, such that the
 nearest-level-spacing distribution has the smallest deviation from the Wigner
 distribution  at $\lambda$ about $1$ in all the models.
 Specifically, $a=0.8$, $1.3$, $0.6$, $1.2 $, and $2$ for the LMG, Dicke,
 defect Ising, defect XXZ, and WBRM models, respectively.
 For brevity, in what follows, we omit the prime in $\lambda'$.

\subsection{Average shape of EFs and their NPT regions} \label{sect-dis-pre}


 Before discussing statistical properties of EFs, it is useful to compute
 the average shape of EFs, which we denote by $\Pi(\varepsilon)$,
\begin{equation}\label{Pi}
 \Pi(\varepsilon)= \la |C_{\alpha k}|^2 \ra_\varepsilon,
\end{equation}
 where $\la \cdot \ra_\varepsilon$ indicates taking average for a given value of $\varepsilon$
 for $\varepsilon_{\alpha k} = E^0_k - E_\alpha$.
 In numerical computation of $\Pi(\varepsilon)$, we take average over $50$ EFs in
 the middle energy  region in each model.

\begin{figure}
\includegraphics[width=\columnwidth]{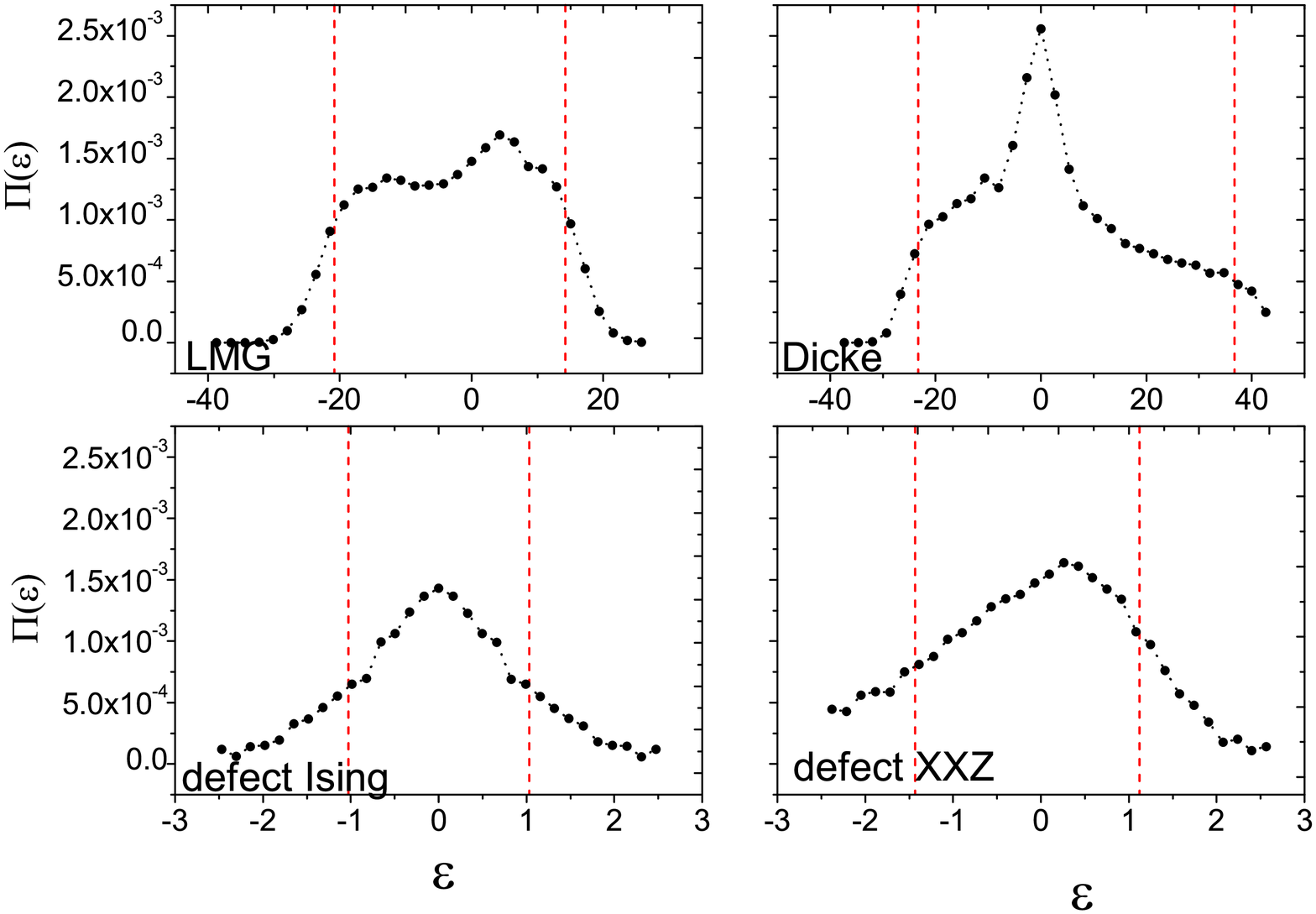}
\caption{ (Color online) Average shape of EFs in the LMG, Dick, defect Ising,
 and defect XXZ models, as a function of the energy difference $\varepsilon$
 [see Eq.(\ref{Pi})].
  Vertical straight lines indicate edges of the averaged NPT regions.
  }
\label{C2n}
\end{figure}

\begin{figure}
\includegraphics[width=\columnwidth]{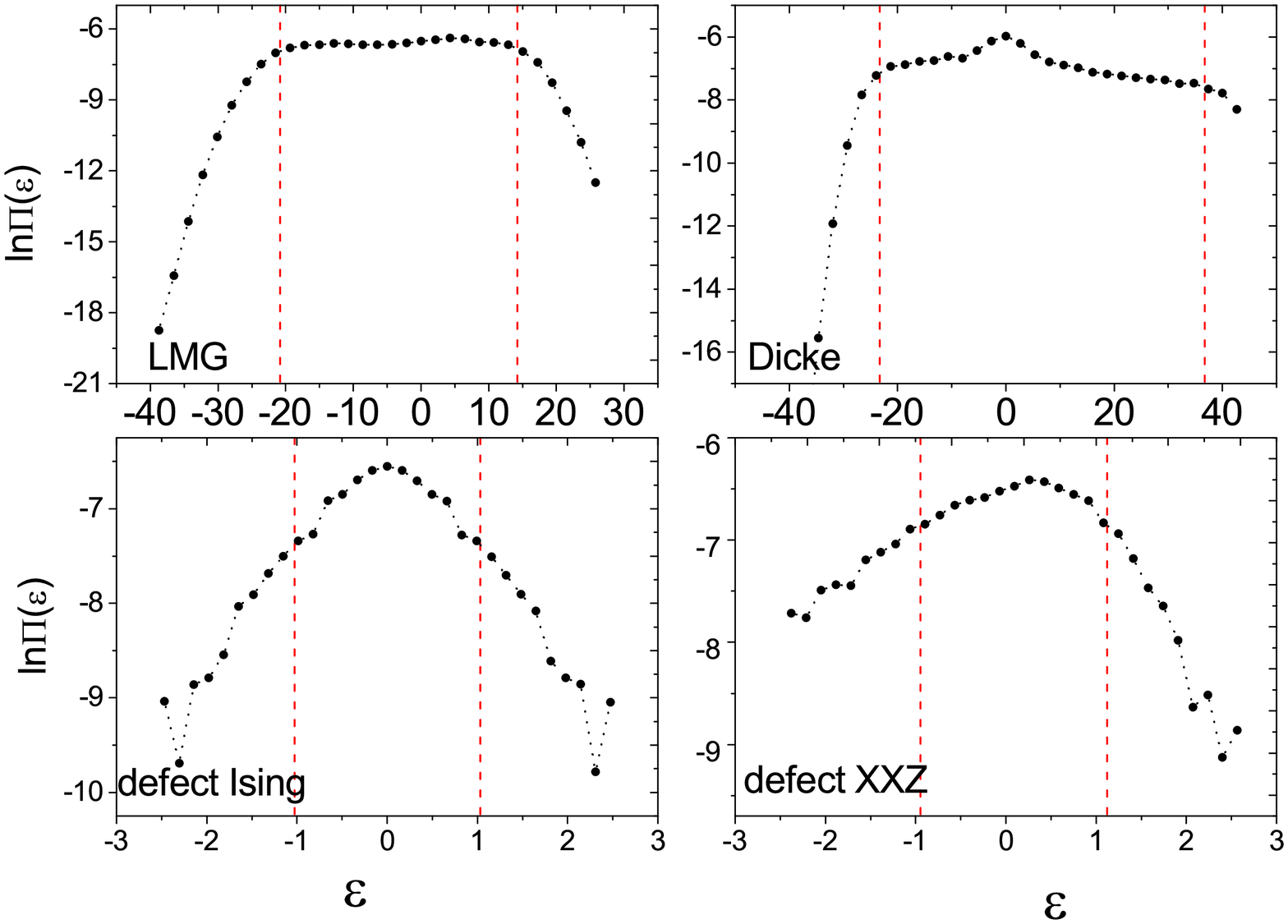}
\caption{
  Same as in Fig.\ref{LC2}, but, in the logarithm scale.
  }
\label{LC2}
\end{figure}

 In models with band structure of the Hamiltonian matrices,
 namely the LMG model, the Dicke model, and the WBRM model,
 as predicted by the GBWPT, main bodies of the EFs indeed lie in the
 NPT-plus-shoulder regions (see Fig.\ref{C2n}).
 Interestingly, even in the defect Ising and defect XXZ models, whose Hamiltonian matrices do not
 have a clear band structure, main bodies of the EFs also lie in the NPT regions.
 As predicted by the GBWPT, in the three models with banded Hamiltonian matrices,
 the EFs behave differently in the NPT and
 PT regions, with quite fast exponential-type decay in the PT regions (see Fig.\ref{LC2})
 (see Ref.\cite{pre00} for numerical results for the WBRM model.)

\section{Distribution of components in models possessing classical counterparts}\label{sect-m-cc}

 In this section, we discuss the distribution of components in NPT parts of EFs in models
 that possess classical counterparts.

\subsection{Distribution in the quantum chaotic regime}

 In models possessing classical counterparts, the matrices of perturbed Hamiltonians
 in unperturbed bases usually have a band structure.
 This is related to the fact that in the semiclassical limit the spectrum of
 the unperturbed system usually does not have an upper bound.
 Meanwhile, the perturbation usually gives a finite contribution to the total energy and,
 as a result, it couples unperturbed basis with a finite energy difference.
 Indeed, both the LMG model and the Dicke model have this property.

 As discussed previously,
 we are mainly interested in statistical properties of the NPT parts.
 Suppose we have $M$ components $C_{\alpha k}$ taken from the NPT parts of considered EFs.
 To compute the distribution of these components,
 we first normalize them, getting $\ww{C}_{\alpha k}$, then, compute the distribution of
 $x=\ww{C}_{\alpha k}\sqrt{M}$,  which we denote by $f(x)$.
 The RMT predicts a Gaussian form of this distribution for the Gaussian orthogonal ensemble (GOE),
 denoted by $f_{\rm GOE}(x)$ \cite{Haake},
\begin{equation}\label{f-GOE}
 f_{\rm GOE}(x)=\frac{1}{\sqrt{2\pi}}\exp(-x^{2}/2).
\end{equation}
 In our numerical computations of the distribution $f(x)$,
 we use the $300$ EFs in the middle energy region in each system.

\begin{figure}
\includegraphics[width=\columnwidth]{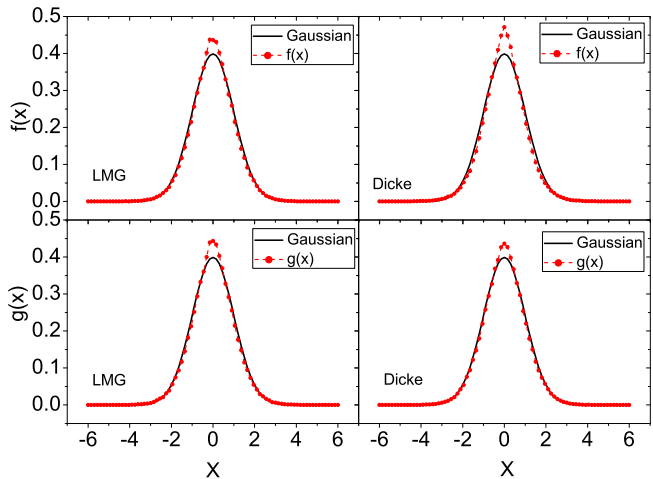}
\caption{(Color online) Upper panels: The distribution $f(x)$ (circles in red) for components
  in the NPT parts of EFs in the middle energy region of the LMG model and the Dicke model at
 $\lambda =1$.
 The solid curve (in black) indicates the Gaussian distribution predicted by the
 RMT in Eq.(\ref{f-GOE}).
 Lower panels: The distribution $g(x)$ for rescaled components in the NPT parts of EFs.
 }\label{dischaosLD}
\end{figure}

 As discussed previously, in the quantum chaotic regime of the LMG model,
 restricted to those EFs that are delocalized in their NPT regions,
 the distribution $f(x)$ is quite close to the Gaussian form $f_{\rm GOE}(x)$ \cite{pre02-LMG}.
 Similar phenomenon has also been observed in the Dicke model.
 However, when all the EFs in the middle energy region are taken into account,
 the distribution $f(x)$ shows some notable deviation from the prediction of GOE
 (see upper panels in Fig.\ref{dischaosLD}), though the deviation is almost negligible between the
 nearest-level-spacing distribution and the prediction of RMT under the same parameters.

 A fact that should be taken into account is that the average shape of EFs,
 i.e., $\Pi(\varepsilon)$, is usually not flat in the NPT regions.
 For example, in the Dicke model, $\Pi(\varepsilon)$ has a high peak (Fig.\ref{C2n}).
 Such a high peak may have non-negligible influence in the distribution of components.
 To take out influence of such peaks, a procedure of `rescaling' can be performed,
 flattening the average shape of the EFs, which has some similarity to the procedure of `unfolding'
 employed in the study of statistical properties of spectra.
 Specifically, we consider the following `rescaled' components,
\begin{equation}\label{}
 \ww C_{\alpha k}=C_{\alpha k}/ \sqrt{ \Pi(\varepsilon)},
\end{equation}
 and use $g(x)$ to indicate the distribution of $\ww C_{\alpha k}$.
 Indeed, in the Dicke model the distribution $g(x)$ becomes closer to $f_{\rm GOE}(x)$
 than $f(x)$, though it still has some small, but notable deviation from $f_{\rm GOE}(x)$.
 (see lower panel in Fig.\ref{dischaosLD}).
 There is almost no change in the LMG model.

 The above-discussed deviation of $g(x)$ from $f_{\rm GOE}(x)$ can be understood
 from a semiclassical point of view.
 As is known, specific dynamics of the underlying classical counterparts,
 such as closed orbits that may lead to scars \cite{KpHl} and small regular islands in the
 chaotic sea, may have non-negligible influences in EFs.
 Such influences can not be moved out by the procedure of `rescaling'.
 Consistently, as to be shown later, the distribution of components can be quite close to
 the prediction of RMT in models not possessing classical counterpart.

\subsection{The process from nearly-integrable to chaotic}

\begin{figure}
\includegraphics[width=\columnwidth]{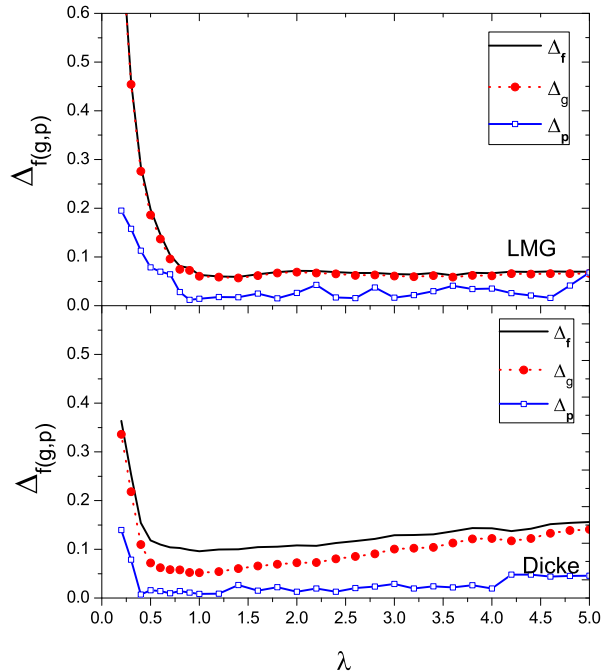}
\caption{Variation of the deviations
$\Delta_f$, $\Delta_g$, and $\Delta_p$ with the perturbation strength $\lambda$
in the LMG and Dicke models.
 }
\label{psdisLDW}
\end{figure}

 In this subsection, we study whether the deviation of $f(x)$ from $f_{\rm GOE}(x)$
 can be employed to characterize the process from nearly-integrable to chaotic.
 In particular, it is of interest to know whether this deviation may give
 information that is not supplied by the spectra.
 Quantitatively, we use the following quantity as a measure of the deviation, i.e.,
\begin{equation}\label{}
\Delta_f=\int \left| f(x)-f_{\rm GOE}(x) \right| dx.
\end{equation}
 Similarly, for the distribution $g(x)$, we use
\begin{equation}\label{}
 \Delta_g =\int \left| g(x)-f_{\rm GOE}(x) \right| dx.
\end{equation}

\begin{figure}
\includegraphics[width=\columnwidth]{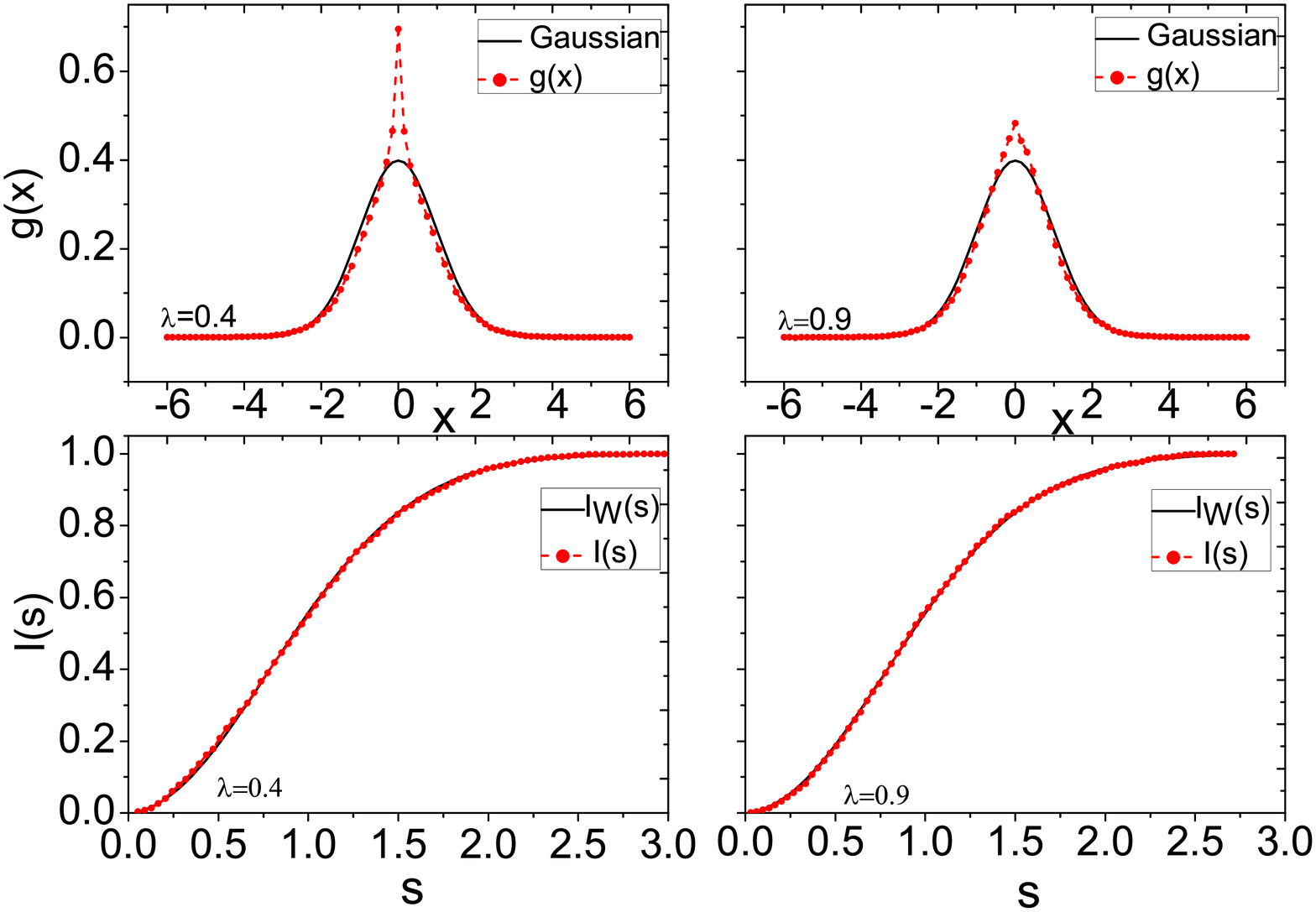}
\caption{ Upper panels: Similar to Fig.\ref{dischaosLD}, for $g(x)$ in the Dicke model at
 $\lambda =0.4$ and $0.9$.
 Lower panels: The corresponding cumulative distribution $I(s)$ (solid circles in read)
 and $I_W(s)$ given by the Wigner distribution (solid curve) in Eq.(\ref{f-Iw}).
}\label{psDicke}
\end{figure}

 For the sake of comparison, we also study variation of the near-level-spacing
 distribution $p(s)$.
 Since the distribution $p(s)$ often has large fluctuations,
 one can instead consider the cumulative distribution of $p(s)$, denoted by $I(s)$,
 $I(s) = \int_{-\infty}^s p(s')ds'$.
 Specifically, we study the quantity
\begin{equation}\label{}
\Delta_p=\int |I(s)-I_{\rm W}(s)| ds.
\end{equation}
 Here, $I_W(s)$ is the cumulative distribution of the Wigner distribution $p_W(s)$,
 which is almost identical to the prediction of RMT,
\be
 p_{W}(s)=\frac{\pi s}{2}\exp\left(-\frac{\pi s^{2}}{4}\right).
\ee
 It is straightforward to verify that
\be\label{f-Iw}
I_{W}(s)=1-\exp \left( -\frac{\pi s^{2}}{4} \right).
\ee
 In each system, the distribution $I(s)$ is computed for the 
 same $300$ EFs in the middle energy region as those used in the computation of $f(x)$.
 To improve the statistics, for each value of $\lambda$, we also use
 $10$ values of $\lambda'$ within the interval $[\lambda-\delta, \lambda +\delta]$
 with $\delta=0.05$.

 Variation of the deviations $\Delta_f$, $\Delta_g$ and $\Delta_p$ with the perturbation
 strength $\lambda$ are given in Fig.\ref{psdisLDW} for the two models.
 Although as discussed above the two distributions $f(x)$ and $g(x)$
 always have some notable deviation from
 $f_{\rm GOE}(x)$ even in the chaotic regime, their variations show a trend similar
 to that of $\Delta_p$.
 Hence, these two distributions are also useful in characterizing
 the process from nearly-integrable to chaotic.
 In the LMG model, we found that,
 similar to the case of chaotic regime discussed above,
 the `rescaling' procedure introduces quite small modification to the distribution of components
 and, as a result, the values of $\Delta_g$ are close to $\Delta_f$ in the whole parameter regime.
 While, in the Dicke mode, $\Delta_g$ are obviously smaller than $\Delta_f$.

\begin{figure}
\includegraphics[width=\columnwidth]{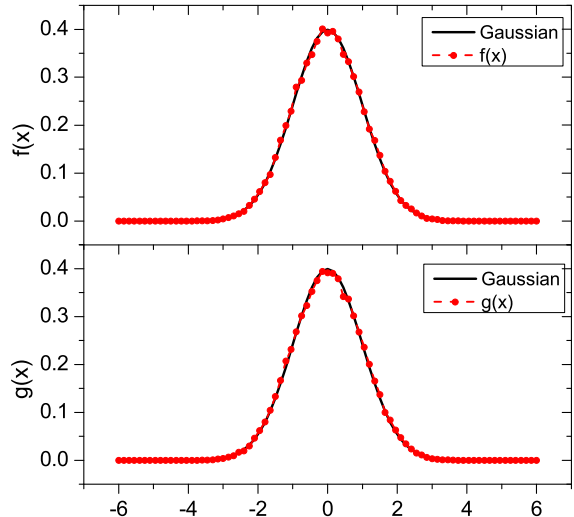}
\caption{(Color online) Similar to Fig.\ref{dischaosLD}, but for the WBRM model at $\lambda=1$.
 }\label{disWBRM}
\end{figure}

 In the process of approaching the chaotic regime in the Dicke model,
 there is some discrepancy between the behaviors of
 $\Delta_f$ (similar for $\Delta_g$) and those of $\Delta_p$ (Fig.\ref{psdisLDW}).
 To be specific, let us use $\lambda^f_m$ and $\lambda^p_m$ to denote the values of $\lambda$
 at which $\Delta_f$ and $\Delta_p$ first reach their (approximate) minimum
 values, respectively.
 In the Dicke model, $\lambda^f_m \approx 0.9 > \lambda^p_m \approx 0.4$.
 This implies that there exists a parameter regime, namely $(\lambda^p_m,\lambda^f_m)$,
 in which the nearest-level-spacing distribution $p(s)$ has already become
 close to the Wigner distribution,
 while the distribution $f(x)$ still has obvious deviation from
 the closest form it may have to $f_{\rm GOE}(x)$.
 We give some examples of this phenomenon in Fig.\ref{psDicke},
 showing $g(x)$ and $I(s)$ at $\lambda=0.4$ and $0.9$.

 We call the above-discussed interesting phenomenon of $ \lambda^p_m < \lambda^f_m$
 a \emph{delay effect} of the distribution of components.
 In the LMG model,  $\lambda^f_m \approx \lambda^p_m$
 and there is no obvious decay effect.
 The decay effect suggests that the spectrum and the EFs
 do not reach the chaotic regime simultaneously.
 Intuitively, this phenomenon is not quite unexpected.
 In fact, the most significant feature of the spectrum of a quantum chaotic
 system is level repulsion, while, to have level repulsion, it is unnecessary
 for the EFs to reach the most irregular form it may have.
 The delay effect also exists in the three models to be discussed in the next section.

\section{Distribution of components in models without classical counterparts}\label{sect-m-non-cc}

 In this section, we discuss the three models which do not have any classical counterpart.
 We first discuss the WBRM model, whose Hamiltonian matrix has a clear band structure,
 then, discuss the defect Ising and the defect XXZ models, whose Hamiltonian matrices
 do not have a clear band structure.

\subsection{The WBRM model}\label{sect-WBRM}

\begin{figure}
\includegraphics[width=\columnwidth]{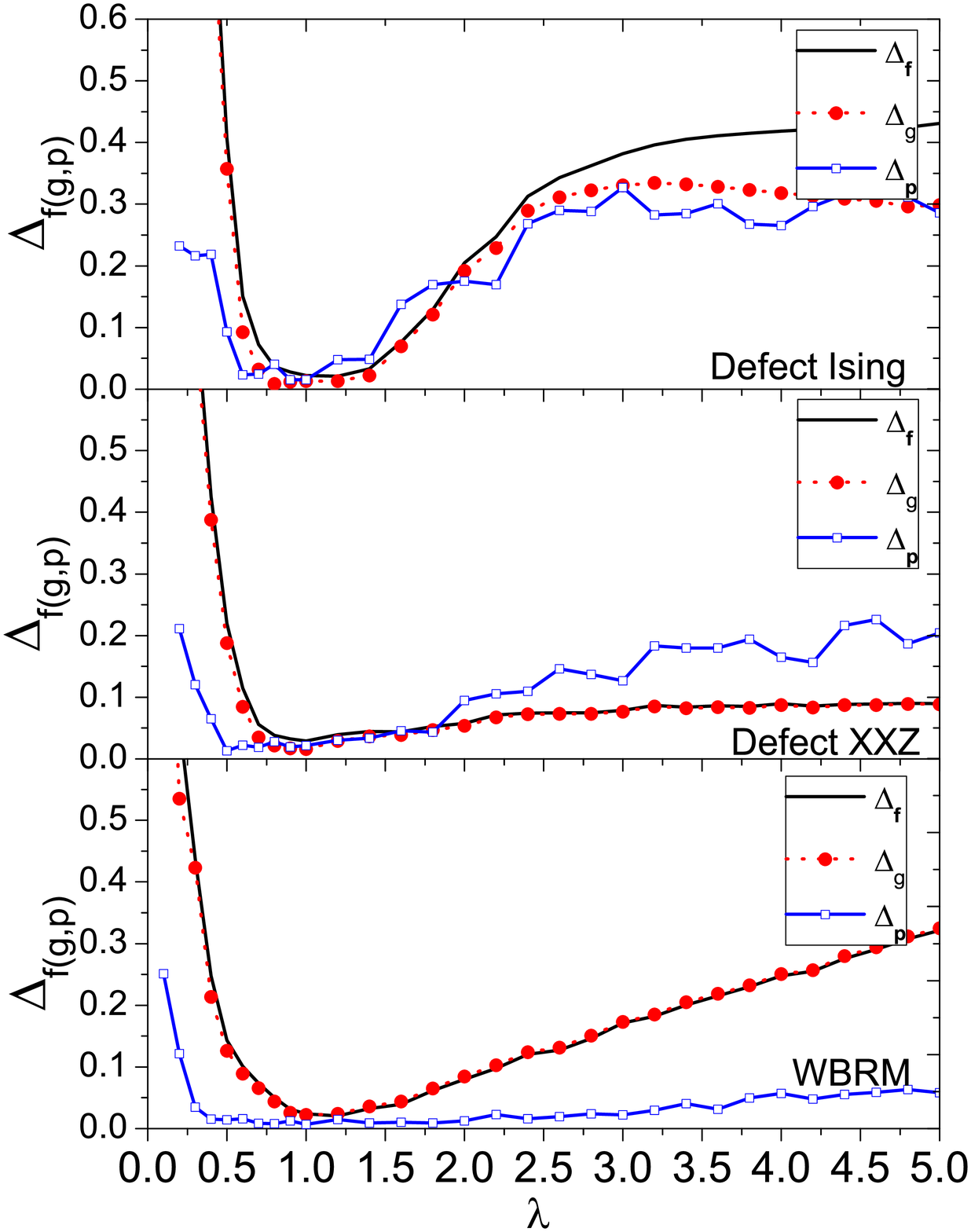}
\caption{ Similar to Fig.\ref{psdisLDW}, but, for the WBRM model, the defect
Ising model, and the defect XXZ model.
}
\label{psdisDIDH}
\end{figure}

 In the WBRM model, when the perturbation is not strong with $\Gamma \ll b$,
 where $\Gamma=2\pi \overline{|\lambda V_{jk}|^{2}} \rho(E) $,
 main bodies of the averaged EFs and local spectral density of states have
 approximately a Lorentz shape with width given by $\Gamma$
 \cite{EF-Flb,EF-Fyodorov,EF-Fyodorov2}.
 Here, $\rho(E)$ is the density of states.
 At relatively strong perturbation with $\Gamma \approx b$, the averaged EFs
 have a Gaussian shape.
 At strong perturbation with very large $\lambda$, this model shows an interesting
 phenomenon called ``localization in energy shell'' \cite{CCGI96}, which is due to
 localization of EFs in their NPT regions \cite{pre00}.

 This model does not have any classical counterpart, hence, unlike in the models of LMG and Dicke,
 there is no influence coming from specific, classical dynamics.
 In fact, in this model the distribution $f(x)$ can be quite close to the prediction of RMT
 (see Fig.\ref{disWBRM} for $\lambda=1$).
 Furthermore, the averaged EF has a relatively flat shape inside the NPT region,
 as a result, the two distributions $f(x)$ and $g(x)$ are close to each other,
 as well as the two deviations $\Delta_f$ and $\Delta_g$ (see Fig.\ref{psdisDIDH}).
 In the computation of the distributions of components,
 $10$ realizations of the perturbation $V$ have been used for each value of $\lambda$.

 As seen in Fig.\ref{psdisDIDH}, the delay effect is quite significant in this model,
 with $\lambda_m^p \approx 0.4$ and $\lambda_m^f \approx 1$.
 In fact, this model has a feature not possessed by the LMG and Dicke models,
 namely, the NPT regions of EFs are quite narrow at relatively small $\lambda$;
 for example, they have only about $5$ components at $\lambda=0.5$.
 The narrowness of the NPT regions of EFs prevents the distribution $f(x)$ from becoming
 quite close to the prediction of RMT.

\begin{figure}
\includegraphics[width=\columnwidth]{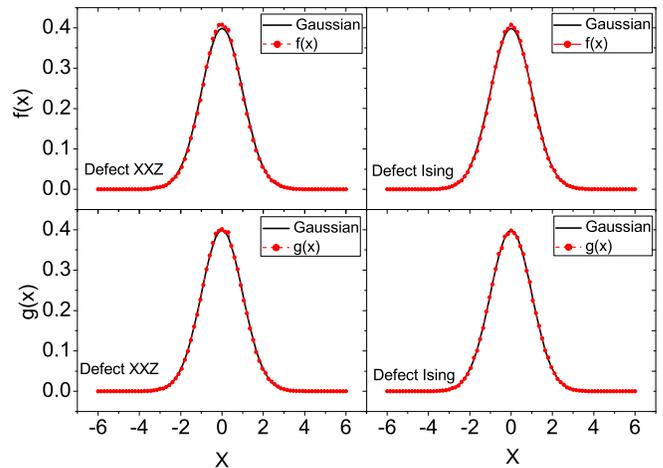}
\caption{(Color online) Similar to Fig.\ref{dischaosLD}, but for the defect Ising
 and defect XXZ models at $\lambda =1$.
 }\label{dischaosDIDH}
\end{figure}

 With increasing $\lambda$, the number of components in the main bodies of the EFs increases.
 At $\lambda=1$, as seen in Fig.\ref{disWBRM}, the distributions $f(x)$ and $g(x)$
 are quite close to the Gaussian distribution.
 However, with further increase of $\lambda$, 
 deviation of $f(x)$ from $f_{\rm GOE}(x)$ increase and
 this is due to the phenomenon of `localization in energy shell' mentioned above.

\subsection{The defect Ising model and defect XXZ model}\label{sect-nonband-H}

 The defect Ising model and the defect XXZ model do not have any classical counterpart, either.
 Similar to a feature discussed above for the WBRM model, without influence from specific
 classical dynamics, the distributions $f(x)$ and $g(x)$ can be quite close to
 the prediction of the RMT  in the quantum chaotic regime
 (see upper panels in Fig.\ref{dischaosDIDH} at $\lambda =1$).

\begin{figure}
\includegraphics[width=\columnwidth]{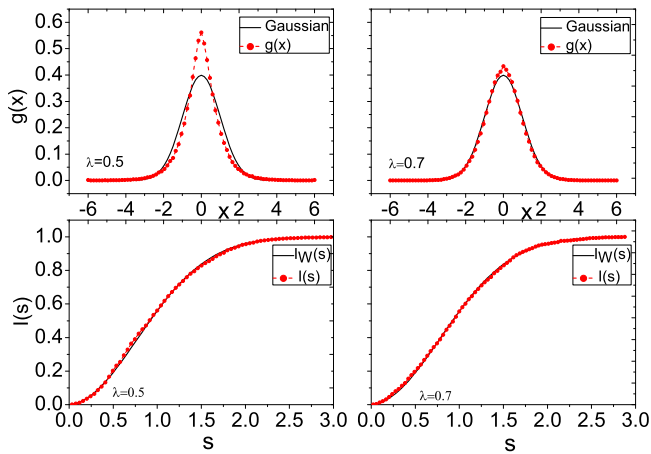}
\caption{(Color online) Similar to Fig.\ref{psDicke}, but for the defect XXZ model at
 $\lambda =0.5$ and $0.7$.
 }\label{disIsing}
\end{figure}

 In the process from nearly-integrable to chaotic, variation of
 the two deviations $\Delta_f$ and $\Delta_g$ show a trend similar to that
 of $\Delta_p$ obtained from the spectra.
 Hence, these two deviations can also be useful in characterizing the process.
 The delay effect also exists in these two models, with
 $\lambda_m^p$ between $0.6$ and $0.7$ and $\lambda_m^f \approx 0.8$ in the defect Ising model,
 and $\lambda_m^p \approx 0.5$ and $\lambda_m^f \approx 0.7$ in the defect XXZ model.
 Some examples of this effect in the defect XXZ model are shown in Fig.\ref{disIsing}.

 Before concluding this section, we discuss briefly the distribution of all components of EFs,
 including both the PT and NPT parts,
 in the quantum chaotic regime in the five models discussed above.
 As expected, the distribution $f(x)$ always shows obvious deviation from the prediction of
 RMT in all the five models.

\begin{figure}
\includegraphics[width=\columnwidth]{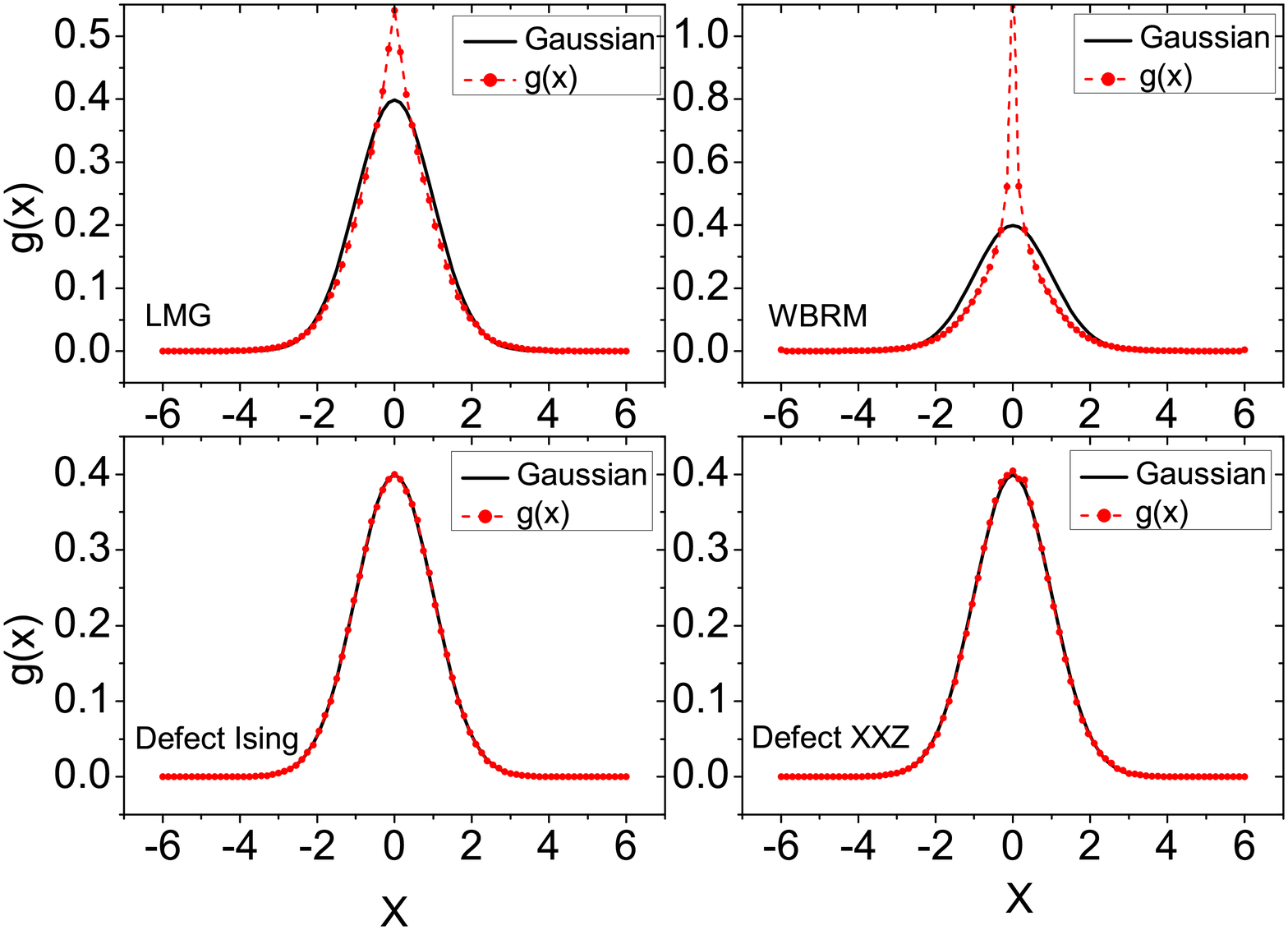}
\caption{(Color online) The distribution $g(x)$ for all (``rescaled'') components of the EFs
 in the LMG, the WBRM, the defect Ising, and the defect XXZ models at $\lambda=1$.}
\label{disall}
\end{figure}

 However, the distribution $g(x)$ of `rescaled' components shows different features, depending on
 whether the Hamiltonian matrices have a clear band structure or not.
 Specifically, in the three models of LMG, Dicke, and WBRM with banded Hamiltonian matrices,
 the distribution $g(x)$ always shows large deviation from the prediction of RMT,
 but, the distribution becomes close to the prediction of RMT 
 in the two models of defect Ising and defect XXZ
 at some values of $\lambda$ (see examples in Fig.\ref{disall}).
 We found that this difference is related to the following observed fact,
 that is, in models with banded Hamiltonian
 matrices, fast-decaying long tails of individual EFs usually have large deviations from
 their averaged shape $\Pi(\varepsilon)$.

\section{Summary and discussions}
\label{sect-Conclusion}

 As well known, statistical properties of the spectra of quantum chaotic systems
 have universal features described by the RMT, only depending on symmetry of the system.
 However, in order to understand statistical properties of EFs of quantum chaotic systems,
 specific dynamics can not be neglected.
 A important problem faced in this study is to make clear what part of EFs in what
 type of systems may show similar statistical behaviors.
 In this paper, by means of numerical simulations performed in five models,
 we study this problem
 in quantum chaotic systems, as well as in the process from nearly-integrable to chaotic.
 Below, we summarize our results and give brief discussions.

 (i) Loosely speaking, main bodies of EFs lie in the NPT (non-perturbative) parts of the EFs.
 In the study of statistical properties of EFs,
 it proves useful to consider their NPT and PT parts separately,
 particularly in models whose Hamiltonian matrices have a clear band structure.
 Below, we discuss the NPT parts, unless otherwise stated.

 (ii) In the process from nearly-integrable to chaotic, deviation of the distribution of
 components in NPT parts of EFs from the prediction of RMT shows a trend
 similar to that given by the statistics of spectra.
 Hence, the distribution of components is also useful in characterizing the process.

 (iii) The distribution of components shows some different features, depending on whether
 the studied model possesses a classical counterpart or not.
 Specifically, in the two models possessing classical counterparts,
 the distribution always shows some notable deviation from the prediction of RMT
 in the quantum chaotic regime, which may be at least partially attributed to
 specific dynamics of the underlying classical system.
 In contrast, in the three models without any classical counterpart,
 the distribution of components can become quite close to the prediction of RMT.

 This phenomenon deserves further investigation.
 By no means should one expect that not-possessing classical counterpart could be a sufficient
 condition for the distribution of components to be close to the prediction of RMT in the
 quantum chaotic regime.

 (iv) The distribution of components may show a delay effect in the process from
 nearly-integrable to chaotic, compared with the statistics of spectra.
 Specifically, in four of the models studied, there exists a regime of perturbation strength,
 in which the distribution of components shows obvious deviation from its closest form
 to the prediction of RMT,
 while the nearest-level-spacing distribution has already become quite close to the prediction of RMT.
 (In the rest model, namely, the LMG model, the width of such a regime can be neglected.)

 Analytical explanation of this phenomenon seems a tough task.
 Anyway, this phenomenon suggests that closeness of the nearest-level-spacing distribution
 to the prediction of RMT may not necessarily imply completely-chaotic motion
 of the EFs.

 (v) In models whose Hamiltonian matrices have a clear band structure,
 the distribution of components in the PT parts of EFs
 always shows large deviation from the prediction of RMT, even after `rescaling'.
 But, in models whose Hamiltonian matrices do not have a clear band structure,
 the distribution of `rescaled' components in the PT parts
 has a form close to the prediction of RMT in the quantum chaotic regime.

 \acknowledgements

 This work was partially supported by the Natural Science Foundation of China under Grant
 Nos.~11275179 and 11535011,
 and the National Key Basic Research Program of China under Grant
 No.~2013CB921800.

\end{document}